\documentclass[letterpaper,twocolumn,10pt]{article}
\usepackage{usenix,epsfig,endnotes}
\usepackage{amssymb}
\usepackage{graphicx}
\usepackage{adjustbox}
\usepackage{subcaption}
\usepackage{multirow}
\usepackage[linesnumbered,ruled,vlined]{algorithm2e}
\usepackage{booktabs} 
\usepackage{amsmath}
\usepackage{booktabs} 
\usepackage{array} 
\usepackage{hyperref}
\usepackage{lipsum}
\usepackage{pdfpages}
\usepackage{svg}
\usepackage{xcolor} 
\usepackage[final]{changes} 
\usepackage{tcolorbox} 

\def\thename{iCHEETAH}

\usepackage{tikz}

\definechangesauthor[name={SY}, color=orange]{SY}
\definechangesauthor[name={chao}, color=red]{chao}

\newcommand{\don}[1]{\textcolor{magenta}{Dongfang: #1}}

\newcommand{\sun}[1]{\textcolor{purple}{[Sun: #1]}}

\begin{document}
\date{}

\title{Enabling Practical and Privacy-Preserving Image Processing}


\author{
{\rm Chao Wang}\\cwang17@wpi.edu\\
Worcester Polytechnic Institute
\and
{\rm Shubing Yang}\\sueyoung@uw.edu\\
University of Washington
\and 
{\rm Xiaoyan Sun}\\xsun7@wpi.edu\\
Worcester Polytechnic Institute
\and 
{\rm Jun Dai}\\jdai@wpi.edu\\
Worcester Polytechnic Institute
\and 
{\rm Dongfang Zhao}\\dzhao@uw.edu\\
University of Washington
}

\maketitle

\thispagestyle{empty}

\renewcommand{\thefootnote}{} 

\footnotetext[1]{Chao Wang and Shubing Yang made equal contributions to this work. Corresponding authors: Drs. Dongfang Zhao and Jun Dai.}

\subsection*{Abstract}

Fully Homomorphic Encryption (FHE) enables computations on encrypted data, preserving confidentiality without the need for decryption. However, FHE is often hindered by significant performance overhead, particularly for high-precision and complex data like images. Due to serious efficiency issues, traditional FHE methods often encrypt images by monolithic data blocks (such as pixel rows), instead of pixels. However, this strategy compromises the advantages of homomorphic operations and disables pixel-level image processing. In this study, we address these challenges by proposing and implementing a pixel-level homomorphic encryption approach, \thename, based on the CKKS scheme.
To enhance the computational efficiency of CKKS in image processing, we propose three novel caching mechanisms to pre-encrypt radix values or frequently occurring pixel values, substantially reducing redundant encryption operations. Extensive experiments demonstrate that our approach achieves up to a 19-fold improvement in encryption speed
compared to the original CKKS, while maintaining high image quality. Additionally, real-world image applications such as mean filtering, brightness enhancement, image matching and watermarking are tested based on FHE, showcasing up to a 91.53\% speed improvement. We also proved that our method is IND-CPA (Indistinguishability under Chosen Plaintext Attack) secure, providing strong encryption security. These results underscore the practicality and efficiency of \thename, marking a significant advancement in privacy-preserving image processing at scale.


\section{Introduction}

Fully homomorphic encryption (FHE), a revolutionary concept pioneered by Gentry~\cite{gentry2009, Gentry2013}, has emerged as a focal point in privacy-preserving computational research. Its unprecedented ability to perform arbitrary computations on encrypted data without decryption has catalyzed a paradigm shift in secure data processing, particularly in sensitive domains such as healthcare and finance. Various FHE schemes have been proposed, including BFV~\cite{FV2012,brakerski2012}, BGV~\cite{BGV2014} and CKKS~\cite{ckks2016}, each designed to address specific computational needs and applications. However, despite its transformative potential, FHE continues to face challenges in computational efficiency and practical implementation when applied to real-world scenarios, especially in the realms of cloud computing and medical imaging~\cite{sathishkumar2024improving,med, reddi2024privacy,medFL,imageclassificaton}, where data integrity and confidentiality are paramount~\cite{chillotti2020tfhe}.

While extant research on homomorphic encryption has predominantly concentrated on optimizing underlying mathematical constructs and enhancing the efficiency of fundamental arithmetic operations on ciphertexts~\cite{kim2022approximate}, the specific challenges inherent in applying FHE to image data have received comparatively little attention.
Due to the substantial computational overhead and low efficiency, current implementations for FHE-based image encryption typically treat images as monolithic data blocks~\cite{cnnfhe}, encrypting them without considering the images' intrinsic structure and properties. An example is to encrypt the image with CKKS by pixel rows, instead of individual pixels, to reduce the number of encryption operations and thus reduce total time needed for encryption. However, this approach compromises the advantages of homomorphic operations and greatly impedes the practical application of FHE in image-related tasks~\cite{son2019revisiting}. 



In the domain of image processing, encryption by pixel is often required to offer unparalleled flexibility and granularity. Despite its higher computational demands, pixel-wise encryption enables a diverse range of sophisticated homomorphic operations directly on encrypted data~\cite{imageCompression}. This method excels in facilitating advanced image processing techniques such as mean filtering, brightness adjustment, and watermarking, all while maintaining data privacy~\cite{CKKSImage2019}. The pixel-level granularity not only enhances privacy protection but also provides precise control over each image element, making it ideal for applications dealing with sensitive visual data or requiring intricate manipulation of encrypted images.
While encryption by row or even by image offers certain performance and efficiency advantages, particularly in processing speed and resource utilization, encryption by pixel presents a more versatile foundation for advanced homomorphic image processing. Although it may initially seem less efficient due to the increased number of encryption operations, the pixel-wise approach opens up a wealth of possibilities for complex algorithms and fine-grained data manipulation. 



In response to these limitations, we propose a novel pixel-level homomorphic encryption approach, iCHEETAH (image-focused Caching-optimized fully Homomorphic Encryption for Efficient Transformation of Assets with High resolutions) based on the CKKS framework, which is renowned for its support of approximate arithmetic~\cite{sara2019image}. This granular approach facilitates fine-grained pixel-level control over encrypted data and enables the direct implementation of various pixel-level image processing operations on ciphertexts. 
The reason why we choose CKKS as the base scheme is because CKKS is the only FHE scheme supporting homomorphic operations over floating numbers, which are commonly used in image processing (e.g., intermediate results). 
To combat the efficiency issue of pixel-level encryption, we introduce three types of caching mechanism, radix-based caching, scanning-based caching, and full caching, to pre-compute and store encrypted radix or frequently occurring pixel values. This approach achieves a remarkable acceleration in the encryption process. 
Particularly, the full caching mechanism results in a more than 19-fold speed improvement compared to original CKKS methods.
This substantial performance boost makes large-scale image encryption and processing both feasible and highly efficient, underscoring the practical potential of FHE-based privacy-preserving techniques.




Furthermore, to showcase the versatility of our pixel-level homomorphic encryption approach, we implement a comprehensive suite of fundamental image processing operations, including image brightening, contrast enhancement, and denoising, directly on encrypted data. By leveraging the homomorphic properties of the CKKS scheme, we execute these operations without compromising the encryption of the images, thereby ensuring the utmost privacy and security of sensitive visual information. Our experimental results demonstrate the efficacy of our approach and highlight its potential to enable privacy-preserving image processing across various domains. \par

\textbf{Threat Model}. 
Our threat model involves a secure system for storing and processing confidential images (such as medical ones), utilizing pixel-level homomorphic encryption.
We assumes that users (such as medical institutions) need to upload sensitive images to the central storage (such as cloud, or central servers) for processing and sharing. The primary assets we aim to protect are the images and the associated encryption keys.
The adversary in our model is assumed to have potential access to the central storage, either as an external attacker or a curious but honest insider within the organization. We assume the storage provider is ``honest but curious''\, correctly executing operations but potentially attempting to learn about the data they're processing~\cite{XIONG201953,threatmodeling}. The users and institutions involved are considered trusted entities with secure local environments for encryption and decryption processes.\par

Additionally, our threat model explicitly assumes that while attackers may observe the encrypted data (ciphertext) within the central storage, they are unable to compromise or gain access to the users' encryption keys required to decrypt this data. As a result, even if the attacker has full visibility of the ciphertext, the lack of access to the decryption keys ensures that the data remains secure and unreadable.

\begin{figure}[htbp]
    \centering
    \includegraphics[width=0.5\textwidth]{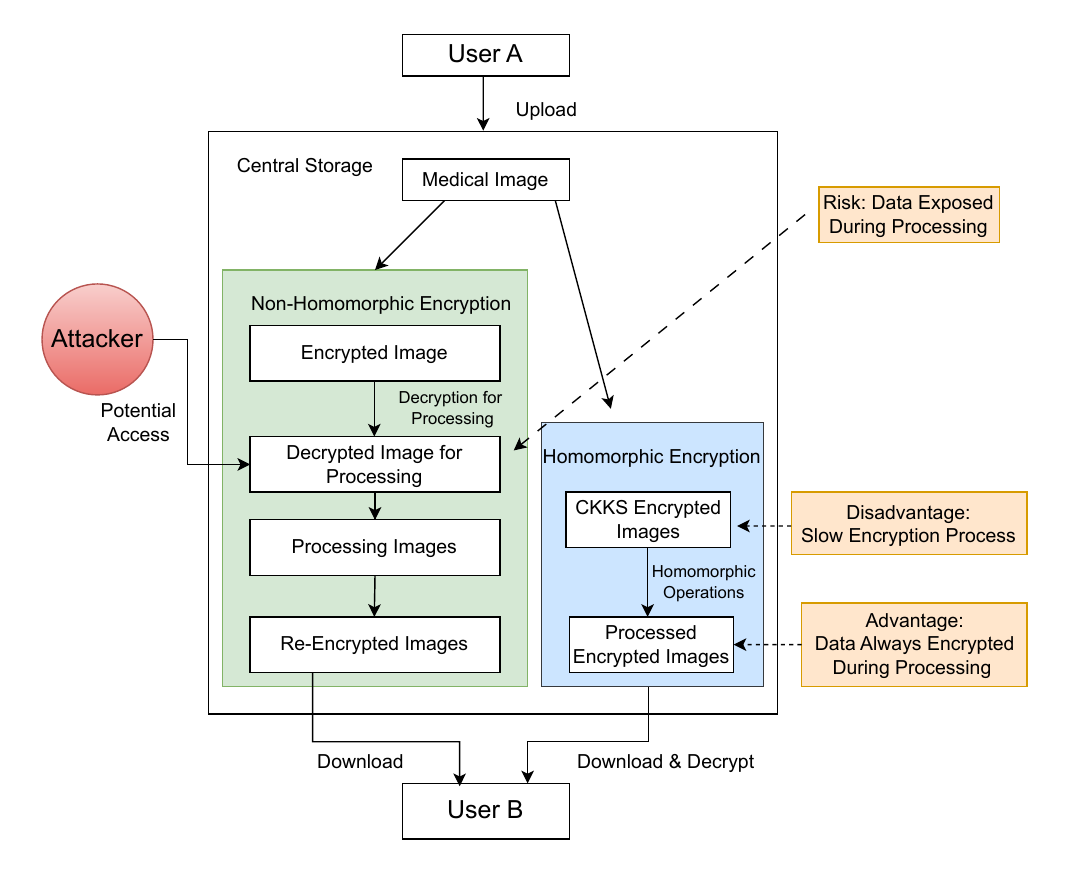}
    \caption{Threat model}
    \label{fig:threatmodel}
\end{figure}

Figure \ref{fig:threatmodel} represents our threat model. It compares two scenarios: non-homomorphic encryption and homomorphic encryption using the CKKS scheme. In the non-homomorphic scenario, images are encrypted before upload but must be decrypted for processing, creating a vulnerability window where an attacker could potentially access sensitive patient data. Our proposed homomorphic encryption scenario uses CKKS to encrypt images before upload, allowing all processing to occur on encrypted data without decryption in the central storage~\cite{GUPTA2020406}. This eliminates the vulnerability window present in the non-homomorphic approach but introduces a trade-off: the initial CKKS encryption process is significantly slower than non-homomorphic methods. The primary security goal is to maintain the confidentiality of images throughout their lifecycle in the central storage, enabling secure processing without exposing unencrypted data.

Our approach guards against potential attacks such as data interception during upload or download and unauthorized access to decrypted images (in the non-homomorphic scenario). The use of CKKS homomorphic encryption mitigates many of these risks by ensuring data remains encrypted during processing. However, the performance impact of CKKS encryption creates a trade-off between security and efficiency that must be carefully balanced. Our work focuses on enhancing the encryption efficiency for pixel-level CKKS-based FHE to make it more practical for real-world image applications, without compromising security.



\textbf{Contributions}. The key insights and contributions of this paper are summarized as follows.
    \begin{itemize}
    
    \item First, we propose a novel pixel-level homomorphic encryption approach that adapts CKKS to operate on individual pixels, enabling fine-grained control over encrypted image data while preserving homomorphic properties (Section \ref{overview}). To dramatically increase processing speed, we introduce three caching mechanisms -- radix-based, scanning-based, and full caching -- that significantly boost the efficiency of CKKS-based encryption, making FHE practical for fast and secure image processing (Section \ref{cachemethodology}).

    \item Second, we prove that our method is IND-CPA (Indistinguishability under Chosen Plaintext Attack) secure, providing strong encryption security. We include a detailed analysis and proof to confirm this, illustrating that our approach combines both high efficiency and robust protection, ensuring it is both secure and practical (Section \ref{INDCPA}).


        
        \item Third, we demonstrate the versatility and practicality of our approach by implementing a range of fundamental image processing operations directly on encrypted data, such as mean filtering, brightness enhancement, image matching, and watermarking. Our experimental results validate the effectiveness of our technique and highlight its potential for enabling privacy-preserving image processing across various domains, including healthcare and finance (Section \ref{casestudy}). We also utilize hundreds of real and complex images from the USC-SIPI~\cite{USCSIPI} database to demonstrate the applicability of our approach to a wide range of images (Section \ref{generalizability}).
    \end{itemize}

\section{Background}
In this section, we briefly review FHE and CKKS, to facilitate a comprehensive understanding.

\subsection{Fully Homomorphic Encryption (FHE)}
Fully Homomorphic Encryption (FHE) stands as a groundbreaking cryptographic method, enabling computations on encrypted data without the need for prior decryption, as initially proposed by Rivest in 1978~\cite{Rivest1978}. Gentry ~\cite{gentry2009} introduced the first viable framework for fully homomorphic encryption using lattice-based cryptography, a type of cryptographic approach that relies on the hardness of mathematical problems related to lattices in high-dimensional spaces. Lattice is a grid-like structure in mathmatics. Gentry's method allows for both addition and multiplication operations on encrypted data, enabling the construction of circuits (such as circuits composed of gates) to perform arbitrary computations~\cite{sota2017,acar2018survey}. The approach \added[id=SY]{of Gentry} involves several stages where noise is initially introduced and then carefully managed~\cite{yi2014homomorphic}. \added[id=SY]{The noise refers to a small amount of random data added to the ciphertext to ensure security. It is introduced as part of the encryption process to make it difficult for an attacker to recover the original plaintext from the ciphertext~\cite{gentry2009}.}This ensures that an unlimited number of additions and multiplications can be performed without excessively amplifying the noise. C. Gentry, A. Sahai, and B. Waters (GSW)~\cite{Gentry2013} in 2013 introduced a new method for constructing FHE schemes that eliminates the costly ``relinearization'' step in homomorphic multiplications.\par



Noise management is critical in FHE. The noise level \( \eta \) in a ciphertext \( c \) is controlled to remain manageable after a series of operations, often using techniques like bootstrapping\added[id=SY]{, which reduces accumulated noise and enables further computations by refreshing the ciphertext through homomorphic decryption and re-encryption, ensuring the ciphertext's integrity remains uncompromised.}



\subsection{CKKS}
\label{ckks}
CKKS (Cheon-Kim-Kim-Song) was originally introduced by Cheon, Kim, Kim and Song in 2016~\cite{ckks2016}. It leverages the RLWE (Ring Learning with Errors) problem, a foundational problem in lattice-based cryptography \added[id=SY]{that involves adding small random errors to polynomial equations, making it difficult for adversaries to solve and ensuring security while supporting efficient homomorphic operations on encrypted data.}  Table \ref{table:parameters} shows key parameters of CKKS we use in this paper~\cite{bts2022}. 

\begin{table}[h]
\resizebox{ 0.45\textwidth}{!}{
\centering
\begin{tabular}{c l}
\hline
\textbf{Symbol} & \textbf{Definition} \\ \hline
$N$ & The ring dimension\\ 
$Q$ & The ciphertext modulus\\ 
$L$ & Maximum level (multiplicative)  \\ 
$\ell$ & Current (multiplicative) level of a ciphertext \\ 
$\text{evk}_{\text{mul}}$ & Evaluation key (evk) for Multiplications \\ 
$\Delta$ &Scaling factor of a CKKS plaintext\\
$\lambda$ & Security parameter of a given CKKS instance \\ \hline
\end{tabular}
}
\caption{Parameters and their definitions in the CKKS scheme}
\label{table:parameters}
\end{table}

CKKS scheme supports arithmetic operations over $\mathbb{C}^{N/2}$. \added[id=SY]{$\mathbb{C}^{N/2}$  represents a space of complex vectors of length $N/2$, where $N$ is a power of 2.}
The plaintext space and ciphertext space are defined over the same domain, as given by the formula $\mathbb{Z}_Q[X]/(X^N + 1)$, 
\added[id=SY]{where $\mathbb{Z}_Q$ is the ring of integers modulo $Q$ and $X$ is the polynomial variable.}

The batch encoding of this scheme maps an array of complex numbers to a polynomial with the property $\mathbb{C}^{N/2} \leftrightarrow \mathbb{Z}_Q[X]/(X^N + 1)$,
where $\leftrightarrow$ denotes the encoding and decoding processes between the two sides.

The property of the encoding and decoding is given by $\text{decode}(\text{encode}(m_1) \otimes \text{encode}(m_2)) \approx m_1 \odot m_2$,
where $\otimes$ represents a component-wise product, and $\odot$ represents a nega-cyclic convolution.

In CKKS, \(N\) represents the ring dimension, and \(Q\) is the ciphertext modulus. Increasing \(N\) and \(Q\) enhances security but also increases computational complexity. A larger \(N\) boosts security by expanding key size and encrypted data complexity. However, it also increases encryption/decryption time and storage needs. Additionally, a larger \(N\) improves noise tolerance, supporting more homomorphic operations. Conversely, a larger \(Q\) enhances precision and noise management but raises computational demands. Thus, choosing \(N\) balances security and performance, while selecting \(Q\) manages precision and noise tolerance.


\subsubsection{RLWE}

RLWE (Ring Learning with Errors)~\cite{rlwe2010,rlwe2015, rlwe2021} 
 is defined over a polynomial ring $R = \mathbb{Z}[X]/f(X)$,
where $f(X)$ is typically a cyclotomic polynomial, and a modulus $q \in \mathbb{Z}$ \added[id=SY]{($q$ is a positive integer that defines the modulus for the coefficients of polynomials)}.

Let $R_q = R/qR$ be the quotient ring. 
Given a secret element $s \in R_q$ and access to a set of noisy ring products, the RLWE problem can be described as $(a_i, b_i = a_i \cdot s + e_i) \in R_q \times R_q$,
where $a_i$ represents uniformly random elements in $R_q$, \added[id=SY]{$b_i$ is the polynomial generated in the encryption process} and $e_i$ denotes small error terms sampled from a discrete Gaussian distribution. The RLWE problem is to distinguish the pairs $(a_i, b_i)$ from uniformly random pairs in $R_q \times R_q$~\cite{rlwe2010, rlwe2015, rlwe2021}.

\subsubsection{CKKS Basics}

RLWE provides a strong foundation for lattice-based cryptography, enabling secure and efficient encryption schemes. CKKS builds on this by extending these capabilities to homomorphic encryption, allowing for approximate arithmetic on encrypted data.

Key generation in CKKS involves the following steps:
\begin{enumerate}
    \item Select a secret key $s \in R_Q$, typically sampled from a small discrete distribution.
    \item Generate a public key $(a, b = a \cdot s + e)$, where $a$ is uniformly random in $R_Q$ and $e$ is sampled from a discrete Gaussian distribution. The public key is $(a, b) \in R_Q \times R_Q$.
    \item Generate an evaluation key required for homomorphic operations.
\end{enumerate}

To encrypt a plaintext $m \in R_Q$, the CKKS encryption process is as follows:
\begin{enumerate}
    \item Encode the plaintext message $m$ into a polynomial $m(X) \in R_Q$, where a cyclotomic polynomial ring is $R_Q = \mathbb{Z}_Q[X]/(X^N + 1)$.
    \item Choose a random polynomial $r \in R_Q$ and compute the ciphertext $(ct_0, ct_1)$ as $ct_0 = r \cdot a + m + e_0$ and $ct_1 = r \cdot b + e_1$,
    where $e_0, e_1$ are small error terms.
\end{enumerate}


To decrypt a ciphertext $(ct_0, ct_1)$, the decryption algorithm computes $m' = ct_0 - s \cdot ct_1 = m + e'$,
where $e'$ is the combined error term. If the error is sufficiently small, the original message $m$ can be recovered by rounding.~\cite{HECO}


\subsubsection{Operations of CKKS}

The CKKS scheme allows for approximate arithmetic on encrypted data, making it well-suited for applications in machine learning and signal processing~\cite{akherati2023low}. In CKKS, plaintexts are encoded as complex vectors, and ciphertexts are structured to support homomorphic addition and multiplication, albeit with some approximation error. There are two basic homomorphic operations in CKKS: homomorphic addition (\(\oplus\)) and homomorphic multiplication (\(\otimes\)).

Homomorphic addition in CKKS is straightforward. Given two ciphertexts $ct_1$ and $ct_2$ that encrypt plaintexts $m_1$ and $m_2$, respectively, their sum is computed as $ct_{add} = ct_1 \oplus ct_2$.
This operation results in a new ciphertext $ct_{add}$ that encrypts the sum of the plaintexts, i.e., $m_1 + m_2$. The addition of ciphertexts does not introduce additional noise beyond what is present in the original ciphertexts, preserving the integrity of the computation.

Homomorphic multiplication is more complex due to the need to handle the growth of noise and manage scaling factors. For two ciphertexts $ct_1$ and $ct_2$, the product is computed as $ct_{mul} = ct_1 \otimes ct_2$.
The resulting ciphertext $ct_{mul}$ encrypts the product of the plaintexts $m_1 \times m_2$. However, multiplication increases both the degree and noise in the ciphertext, making it necessary to apply a process known as relinearization.

Relinearization reduces the degree of the resulting ciphertext back to its original form, which is crucial for maintaining efficiency in further operations. This is done by using an evaluation key $evk$ generated during the key generation phase $ct_{rel} = Relin(ct_{mul}, evk)$.
Here, $ct_{rel}$ is the relinearized ciphertext, which now has a reduced degree, making it more manageable for subsequent operations.

After multiplication, the ciphertext may require rescaling to maintain the balance between the ciphertext and the plaintext. Rescaling involves dividing the ciphertext by a scaling factor $\Delta$ to reduce the noise $ct_{rescaled} = Rescale(ct_{rel}, \Delta)$
This operation helps in maintaining the correctness of the plaintext after decryption and prevents the noise from overwhelming the signal.

\section{Our Approach}
\label{methodology}

In the following sections, we will provide a detailed overview of our approach (Section \ref{overview}), the caching mechanisms (Section \ref{cachemethodology}), how to add randomness (Section \ref{Randomness}), and our approach's compliance to IND-CPA security (Section \ref{INDCPA}). 

\subsection{Overview}
\label{overview}
Our approach is designed to encrypt image data while preserving the homomorphic properties of the ciphertext, thereby enabling ciphertext-based image processing. 
To accomplish this, implementing pixel-level FHE for image encryption is essential, however, it will entail a substantial time overhead. Based on the CKKS scheme, our approach achieves pixel-level image encryption and addresses the incurred time overhead through the strategic use of precomputation and caching. The approach is also structured to ensure that the encrypted data is resistant to chosen plaintext attacks (CPA) while maintaining computational efficiency.

\begin{figure*}[htbp]
    \centering
    \includegraphics[width=\textwidth]{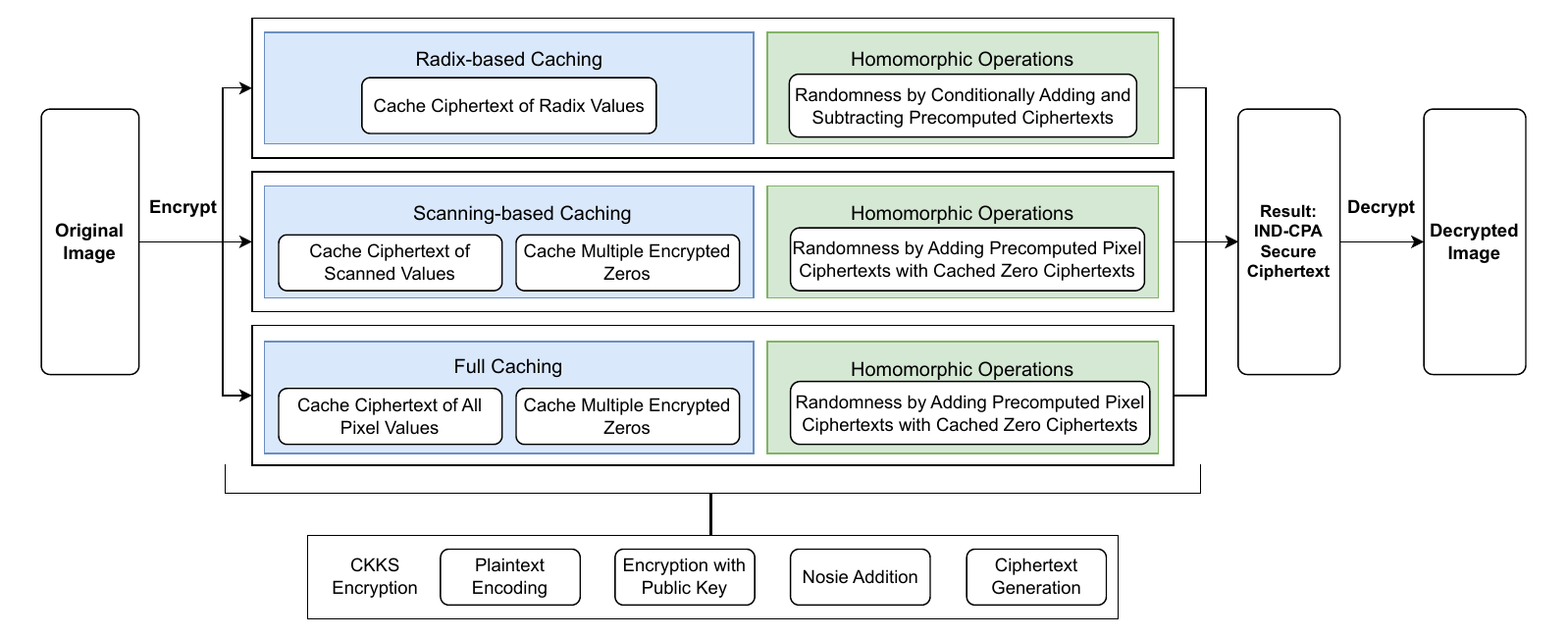}
    \caption{Design of \thename}
    \label{fig:design}
\end{figure*}

Figure \ref{fig:design} shows our approach overview. Adapted from the CKKS scheme, this approach begins with encoding and encrypting plaintext values, which are then cached to be used in subsequent operations. The plaintext values can be encrypted and cached in three ways: Radix-based caching that encrypts radix values, scanning-based caching that scans pixel values contained in an image and then encrypts these values, and full caching that encrypts all possible pixel values. 
Based on the using scenario, one of the three caching strategies can be selected to produce the cached ciphertexts. These ciphertexts are then directly used during the image encryption process and thus effectively reduces the computational complexity.

Once the ciphertexts for radices or pixel values are cached, theoretically the image can be encrypted by directly using the cached values to encrypt each pixel. However, this will suffer from CPA because identical pixel values result in identical ciphertexts due to using cached values. Adversaries can reveal the plaintexts by encrypting chosen plaintexts and comparing the ciphertexts, which is known as CPA. Therefore, our approach also needs to add randomness to the ciphertext of each pixel to make sure same plaintexts will generate different ciphertexts. We impose randomness through two different homomorphic operation strategies: 1) For radix-based caching, we perform some extra adding and subtracting operations to the cached ciphertext of the original pixel, such as adding ciphertext of 32, and then substrcting ciphertext of 16 twice.
2) For scanning-based and full caching, we add cached ciphertexts of zeros to the cached ciphertext of the original pixel. The ciphertext of zero can be different every time when zero is encrypted, and thus brings in randomness.
Both strategies do not change the decrypted value for the original pixel, guaranteeing that the encrypted image can be accurately decrypted and the integrity of the original data is preserved. More details about randomness is in Section~\ref{Randomness}.

\subsection{Radix-based Caching, Scanning-based Caching, and Full Caching}
\label{cachemethodology}

To address the performance challenges of pixel-level FHE encryption, which is crucial for enabling ciphertext-based image processing, a strategic approach is needed to accelerate FHE operations. Section \ref{relatedwork: acceleration} reviews existing work on HE acceleration, but few of these methods are directly applicable to the goals of this paper, or fully consider the unique characteristics of image data. In contrast, the design of \thename ~is informed by our insights on image data, with the underlying rationale detailed below:

\emph{Insight 1: Pixel-level image data consists exclusively of integer values.} Consequently, pixel-level FHE image encryption is inherently encrypting a matrix of integers, with each operation carried out at the level of individual pixel values. To expedite the homomorphic encryption of various integer values, it is effective to cache the encryption of specific unit values and reconstruct the ciphertext for a target value by applying homomorphic additions to the cached ciphertexts of the corresponding unit values that sum to the target value. A natural selection for unit values is the radices. This strategy, rooted in math, is consistent with previous works, such as \cite{gadgetDecomposition} and \cite{sigmod2023}, despite their differing objectives and scenarios. This caching strategy based on CKKS is formalized as follows.

\textbf{Radix-based Caching.} Radix-based caching leverages base-$r$ representation to decompose each value in the dataset into a sum of powers of $r$. For any value $p$ in the dataset, it can be represented as a sum of distinct powers of $r$, where each component is then encrypted using the CKKS scheme:
\[
c_p = \sum_{i=0}^{n} \text{CKKS.Enc}(r^{k_i})
\]
By precomputing and caching the ciphertexts for these powers of $r$, computational overhead during the encryption process is significantly reduced, as retrieving the ciphertext for any pixel value becomes a simple lookup operation. A common choice for the radix is 2, i.e., $r=2$, where the CKKS ciphertexts of 0, 2, 4, 8, 16, \dots ~are precomputed and cached.

\emph{Insight 2: Pixel values frequently recur in image data.} Images often contain frequently occurring values within their pixel datasets. By identifying and caching the ciphertexts of these frequently occurring pixel values, we can further optimize the homomorphic encryption process. Specifically, because many images have large regions with uniform or similar pixel values, this strategy can significantly enhance performance in image encryption. Additionally, it is adaptable to handle a wide range of values beyond image data. We refer to this strategy as scanning-based caching.

\textbf{Scanning-based Caching.}  Scanning-based caching conducts a pre-processing scan of the dataset, and identifies the most common values, $p_1, p_2, \dots, p_k$. For each value $p_j$, we compute and store the corresponding ciphertext:
\[
c_{p_j} = \text{CKKS.Enc}(p_j)
\]
Scanning-based Caching can be used in many circumstances. When encrypting the image, these precomputed ciphertexts are retrieved from the cache, thereby avoiding the need to repeatedly encrypt the same pixel values. This caching strategy reduces the computational burden and enhances the overall efficiency of the encryption process, particularly in images with high redundancy or repeated patterns.

\emph{Insight 3: Pixel values in an image are drawn from a finite set of numbers.} An image is a visual representation made up of a grid of pixels, each corresponding to a specific point. In digital imaging, we assume the use of the RGB color scheme, where each pixel's color is defined by a combination of Red, Green, and Blue. Each color channel is represented by a pixel value ranging from 0 to 255, with 0 indicating no intensity and 255 indicating full intensity. We therefore pursue offline caching of all such values, referring to this strategy as full caching. Other schemes such as HSV (Hue, Saturation, Value) and HSL (Hue, Saturation, Lightness) are not focus of this paper and can be the future work.  

\textbf{Full Caching.} Full caching involves precomputing and caching the ciphertexts for all possible pixel values $p$ in the range $[0, 255]$, ensuring that every potential pixel value has a corresponding cached ciphertext $c_p$:
\[
c_p = \text{CKKS.Enc}(p)
\]
This approach is perfectly suited for image processing and ensures that each pixel value in the image can be encrypted efficiently without the need for on-the-fly computation. While scanning-based caching can handle cases where plaintext values are indefinite, full caching are applicable to cases where all possible plaintext values are already known.

\subsection{Adding Randomness}
\label{Randomness}
When ciphertexts are exposed, they become vulnerable to chosen-plaintext attacks (CPA). Thus, it is essential to introduce randomness each time a cached ciphertext is used.

For radix-based caching, we proposed to iterate through each precomputed ciphertext in \(\text{radixes}[i]\) and randomly decide whether to add it to the main ciphertext \(\text{Ctxt}[i]\), where i is the iteration index. If this addition occurs, the process then subtracts \(\text{radixes}[i-1]\) from \(\text{Ctxt}[i]\) a total of \(r\) times, effectively introducing a controlled form of randomness to the ciphertext modification process:
\[
\text{Ctxt}[i] \oplus \text{radixes}[i] \ominus r \times \text{radixes}[i-1]
\]
where \(\oplus\) represents the homomorphic addition between 2 cipertexts and \(\ominus\) represents the homomorphic subtraction between 2 ciphertexts. For example, if radix r=2 and i=5, the randomness for Ctxt[5] can be done through 
\[
\text{Ctxt}[5] \oplus 2^5 \ominus 2 \times 2^4
\]

For scanning-based caching and full caching, we implement a thread pool-like mechanism, called ``\emph{randomness pool}'', by creating a pool consisting exclusively of precomputed ciphertexts of zeros. This pool ensures a continuous and efficient supply of ciphertexts of zeros, which are retrieved as needed during encryption or homomorphic operations. This pool consists of pre-encrypted zero values $\{c_{z1}, c_{z2}, \dots, c_{zn}\}$, where each $c_{zi}$ is an encrypted zero:
\[
c_{zi} = \text{CKKS.Enc}(0)
\]
During the encryption process, for each pixel value $p$, a randomly selected zero ciphertext $c_{zi}$ from the pool is added to the cached ciphertext:
\[
c_{p, \text{final}} = c_p + c_{zi}
\]
This addition ensures that identical pixel values do not produce identical ciphertexts. By maintaining the certain number of encrypted zeros in the randomness opol, \thename ~is provably IND-CPA secure (see Section~\ref{INDCPA}) while reducing the computational load associated with image encryption.

\subsection{Security Analysis}
\label{INDCPA}

In this section, we establish that our modified CKKS scheme, which includes caching ciphertexts for plaintexts \(0\) to \(255\) and the use of pre-encrypted zeros, maintains IND-CPA (Indistinguishability under Chosen Plaintext Attack) security. This is achieved by reducing the security of the modified scheme to that of the original CKKS scheme.

Assume for the sake of contradiction that the modified CKKS scheme \( \tilde{\Pi} \) is not IND-CPA secure. This would imply the existence of a probabilistic polynomial-time (PPT) adversary \( \mathcal{A} \) capable of distinguishing between the encryptions of two plaintexts \( m_0 \) and \( m_1 \) under \( \tilde{\Pi} \) with a non-negligible advantage:
\[
\left| \Pr[\mathcal{A}(\tilde{\Pi}(m_0)) = 1] - \Pr[\mathcal{A}(\tilde{\Pi}(m_1)) = 1] \right| \geq \epsilon(n),
\]
where \( \epsilon(n) \) is a non-negligible function of the security parameter \( n \).

To derive a contradiction, we construct a PPT adversary \( \mathcal{B} \), which uses \( \mathcal{A} \) as a subroutine to break the IND-CPA security of the original CKKS scheme \( \Pi \). The adversary \( \mathcal{B} \) is provided with a public key \( pk \) and receives two challenge ciphertexts \( c_0 = \Pi(m_0) \) and \( c_1 = \Pi(m_1) \) from the IND-CPA challenger for \( \Pi \). Then, \( \mathcal{B} \) randomly selects a ciphertext \( z \) from the set \( Z = \{\Pi(0)\} \) of pre-encrypted zeros. The adversary \( \mathcal{B} \) computes the following ciphertexts:
\[
\tilde{c}_0 = c_0 + z \quad \text{and} \quad \tilde{c}_1 = c_1 + z
\]
Subsequently, \( \mathcal{B} \) feeds the ciphertexts \( \tilde{c}_0 \) and \( \tilde{c}_1 \) to \( \mathcal{A} \) and outputs \( \mathcal{A} \)'s guess as its own.

Since \( \mathcal{A} \) can distinguish between \( \tilde{c}_0 \) and \( \tilde{c}_1 \) with a non-negligible advantage, \( \mathcal{B} \) would similarly distinguish between \( c_0 \) and \( c_1 \), thereby breaking the IND-CPA security of the original CKKS scheme \( \Pi \). This contradicts the established security of the CKKS scheme, which is proven to be IND-CPA secure under the hardness assumption of the Ring-LWE problem. Hence, no such adversary \( \mathcal{A} \) can exist, and we conclude that the modified CKKS scheme \( \tilde{\Pi} \) is also IND-CPA secure.

It is important to note that the use of zero ciphertexts does not significantly impact noise levels, thus preserving the correctness of decryption. Furthermore, caching encryptions of known plaintexts introduces no additional vulnerabilities, as these ciphertexts are generated using the original CKKS encryption function, which is secure.

\section{Evaluation}


This section will present our evaluation objectives (Section~\ref{objectives}), setup of experiments (Section~\ref{experimentalSetup}), performance comparisons (Section~\ref{comparisonsection}), real-world applications as case studies (Section~\ref{casestudy}), and broader applicability (Section~\ref{generalizability}). Notably, our method demonstrates exceptional scalability and can successfully handle operations on 1024$\times$1024 images -- an achievement not previously reported in the literature~\cite{HECO,CKKSImage2019}.

\subsection{Objectives}
\label{objectives}

We aim to address the following questions through our experimental evaluation:\par
\begin{itemize}
    \item How does our pixel-level homomorphic encryption method with caching optimization perform compared to existing methods in terms of encryption time and scalability? (Section \ref{comparisontime})
    \item When performing secure image processing operations for encryption and decryption, to which extent can our approach preserve image quality while maintaining computational efficiency? (Section~\ref{comparisonquality})
    \item Does our approach effectively enable ciphertext-based image processing in real-world image processing and analysis applications? (Section~\ref{casestudy})
    \item Can our approach be broadly applicable to a diverse range of images? (Section~\ref{generalizability})
\end{itemize}

\subsection{Experimental Setup}
\label{experimentalSetup}

Our implementation is based on the Lattigo library~\cite{lattigo}, an open-source Go library for lattice-based cryptography. We chose Lattigo for its efficient implementation of the CKKS scheme and its support for homomorphic operations. \thename\ consists of roughly 4K lines of code (LOC) of Go. Our solution is integrated with the Go image processing library for handling various image operations. The project is developed using Go 1.20.4 and compiled with the standard Go compiler. For parallelization, we utilize the built-in concurrency features of Go, including goroutines and channels.\par
Our experiments are conducted on a high-performance computing cluster of CloudLab Clemson. Each node is equipped with Two 32-core AMD 7542 at 2.9GHz, 512 GB of RAM, and 2 TB of SSD storage. The operating system is Ubuntu 20.04 LTS.\
We utilize the USC-SIPI Image Database~\cite{USCSIPI} for our experiments. This dataset was chosen for its diverse range of image types and content, which allows us to test our encryption method across various scenarios. Additionally, the USC-SIPI database is free from copyright restrictions, making it ideal for research purposes. The dataset includes a wide variety of images, such as textures, aerial photographs, and standard test images, providing a comprehensive basis for our experiments.\par

As we mentioned in Section \ref{ckks}, increasing \( N \) and \(Q\) can indeed enhance the security of the system, but it also introduces trade-offs in computational complexity and performance. Therefore, in our experiments, we chose \(N\) = $2^{12}$ and \(Q\) = $2^{109}$, making sure a security of 128 bits for a secret key with uniform ternary distribution equation $s \in \{-1, 0, 1\}^N$. These parameters also ensure that our experiments can proceed efficiently within the constraints of our limited resources.

Prior to our experiments, we perform several preprocessing steps to standardize the dataset. First, we conduct image cropping and resizing. As the original dataset contains images of varying resolutions, we create subsets of images at specific resolutions: 8x8, 64x64, 128x128, 256x256, 512x512, and 1024x1024 pixels. This standardization allows us to evaluate our encryption method's performance across different image sizes. Second, we convert all images to BMP (Bitmap Image File) format to ensure a consistent file format, which guarantees compatibility and facilitates accurate performance comparisons~\cite{panda2021principal}. BMP is selected for its pixel-based structure, which is ideal for applications such as medical imaging due to its uncompressed format and high image quality. 



\subsection{Performance Comparison}
\label{comparisonsection}
This subsection presents the results of performance comparison between our proposed approaches and the baseline method (without any caching optimization) for pixel-level encryption across a diverse range of images. \par


\subsubsection{Time and Scalability}
\label{comparisontime}

\begin{figure}[!t]
\centering
\includegraphics[width=\columnwidth]{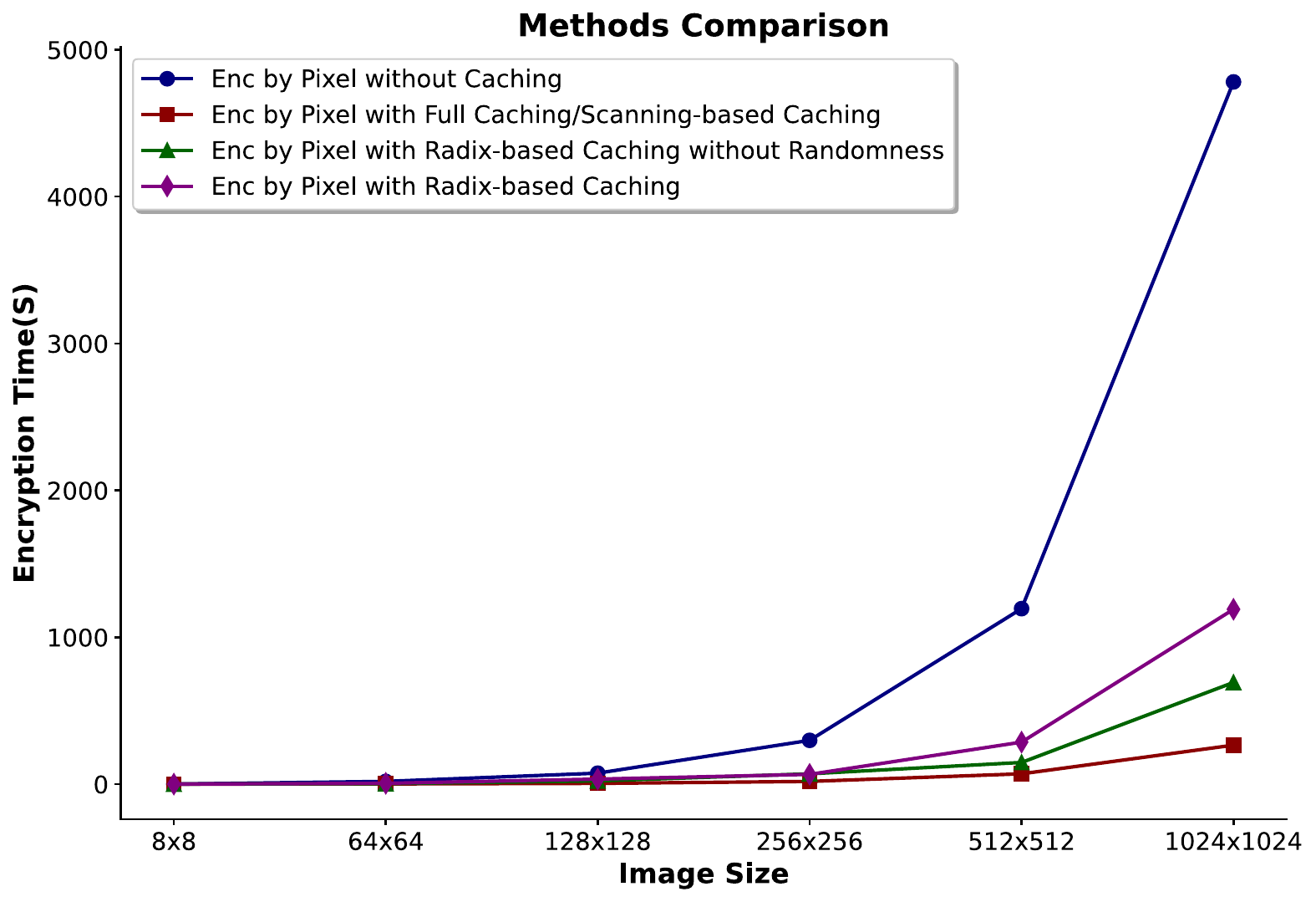}
\caption{Encryption time across different image sizes}
\label{fig: comparison}
\end{figure}

In our evaluation of the encryption methods, we observed significant differences in encryption efficiency. Figure \ref{fig: comparison} compares the encryption times of these methods across various image sizes, presented on a linear scale. The methods analyzed include the baseline pixel-level encryption without caching, pixel-level encryption with scanning-based caching, pixel-level encryption with full caching, and pixel-level encryption using radix-based caching, both with and without randomness.\par


As evident in the graph, the baseline method (Enc by Pixel without Caching) shows a steep increase in encryption time as the image size grows. This method, while straightforward, lacks optimizations to mitigate the computational complexity inherent in pixel-level encryption, particularly for larger images. The encryption time for this method rises exponentially, making it less practical for high-resolution image processing.\par

In contrast, the full caching method (Enc by Pixel with Full Caching) demonstrates a substantial improvement in efficiency across all image sizes. 
The most substantial improvement is observed with the largest image sizes, where the full caching method achieves up to 19 times the performance of the baseline. This method also outperforms the radix-based caching method, regardless of whether randomness is applied to radix-based caching. 

The scanning-based caching, which only differs from full caching by the scanning part which occurs before image encryption, produces nearly identical results to full caching. Consequently, in the graph, the full caching and scanning-based caching methods are represented by the same plot, though the scanning-based caching may encrypt and cache less pixel values than full caching during the caching phase.\par


The radix-based caching method (Enc by Pixel with Radix-based Caching), with or without randomness, exhibits enhanced performance over the baseline, though it does not match the level of efficiency achieved by the full caching method.

The substantial reduction in encryption time observed in our experiments can be attributed to our innovative caching strategy. The full caching method resembles the creation of a lookup table, where the encrypted mapping of each pixel value $p$ is pre-stored as a cached ciphertext. This allows retrieval of the encrypted value in O(1) time, remarkably reducing the computational overhead of encrypting each pixel individually. 
It is important to note that this performance gain significantly outweighs the overhead associated with adding randomness to defeat CPA attacks. Consequently, we transform the traditionally time-consuming encryption process into a swift lookup operation, enabling the efficient processing of large images without compromising security.

\subsubsection{Image Quality}
\label{comparisonquality}

\begin{figure*}
    \centering
    \begin{subfigure}[b]{0.15\textwidth}
        \includegraphics[width=\textwidth]{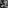}
        
    \end{subfigure}
    \hfill
    \begin{subfigure}[b]{0.15\textwidth}
        \includegraphics[width=\textwidth]{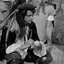}
       
    \end{subfigure}
    \hfill
    \begin{subfigure}[b]{0.15\textwidth}
        \includegraphics[width=\textwidth]{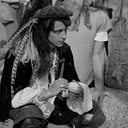}
      
    \end{subfigure}
    \hfill
    \begin{subfigure}[b]{0.15\textwidth}
        \includegraphics[width=\textwidth]{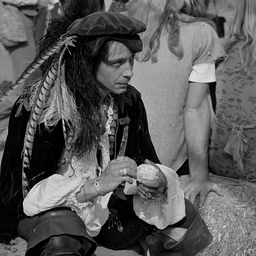}
        
    \end{subfigure}
    \hfill
    \begin{subfigure}[b]{0.15\textwidth}
        \includegraphics[width=\textwidth]{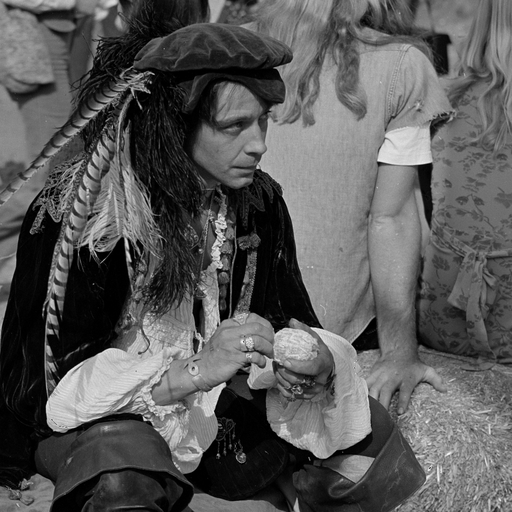}
       
    \end{subfigure}
    \hfill
    \begin{subfigure}[b]{0.15\textwidth}
        \includegraphics[width=\textwidth]{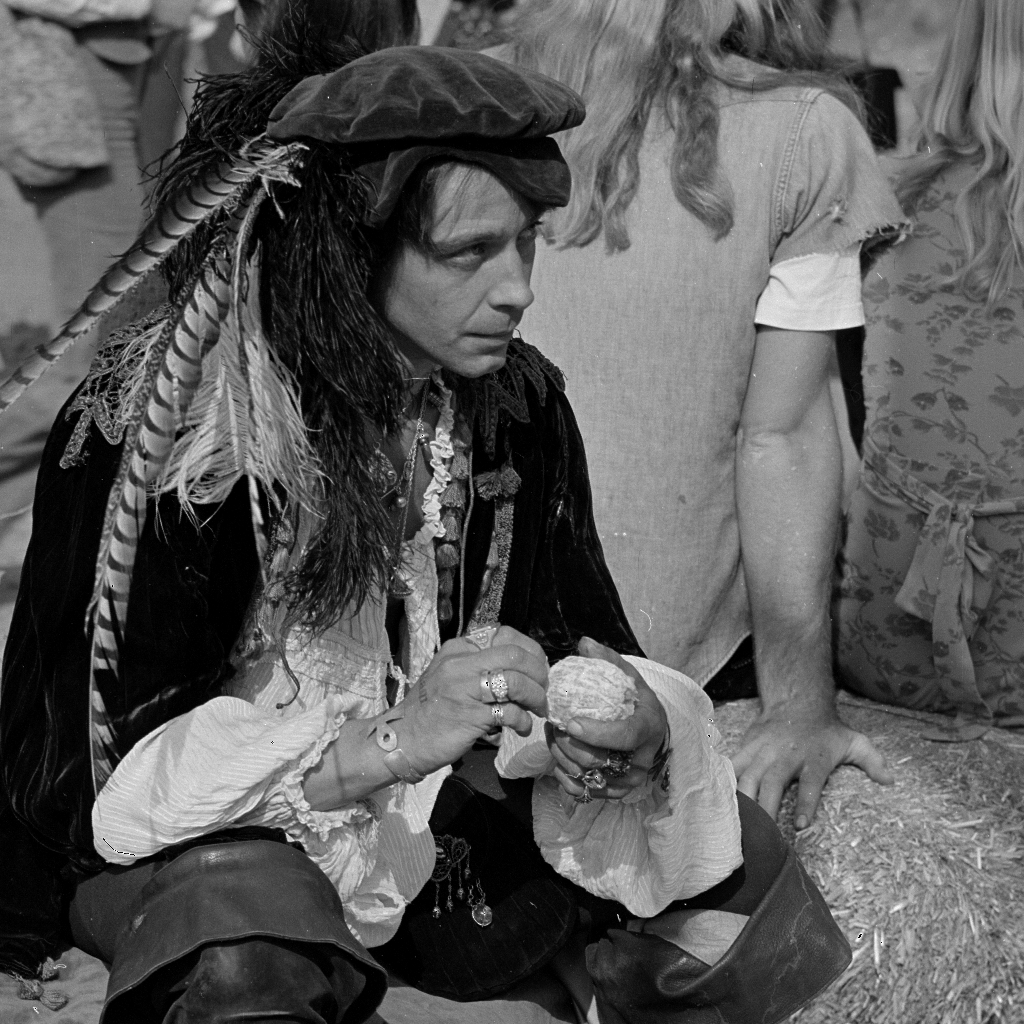}
       
    \end{subfigure}
    
    \vspace{1em}
    
    \begin{subfigure}[b]{0.15\textwidth}
        \includegraphics[width=\textwidth]{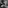}
        \caption{8x8}
    \end{subfigure}
    \hfill
    \begin{subfigure}[b]{0.15\textwidth}
        \includegraphics[width=\textwidth]{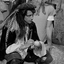}
        \caption{64x64}
    \end{subfigure}
    \hfill
    \begin{subfigure}[b]{0.15\textwidth}
        \includegraphics[width=\textwidth]{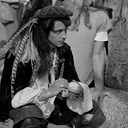}
        \caption{128x128}
    \end{subfigure}
    \hfill
    \begin{subfigure}[b]{0.15\textwidth}
        \includegraphics[width=\textwidth]{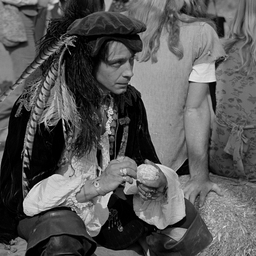}
        \caption{256x256}
    \end{subfigure}
    \hfill
    \begin{subfigure}[b]{0.15\textwidth}
        \includegraphics[width=\textwidth]{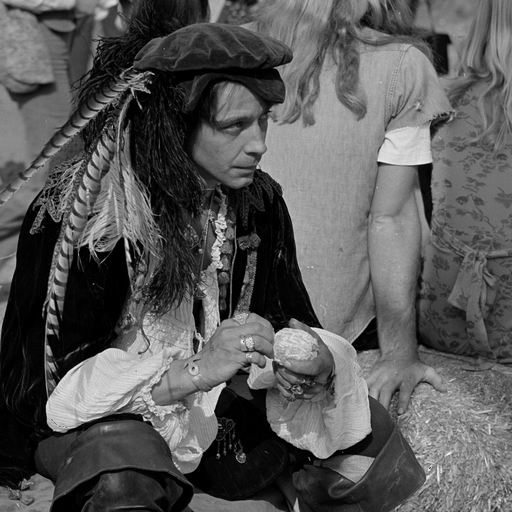}
        \caption{512x512}
    \end{subfigure}
    \hfill
    \begin{subfigure}[b]{0.15\textwidth}
        \includegraphics[width=\textwidth]{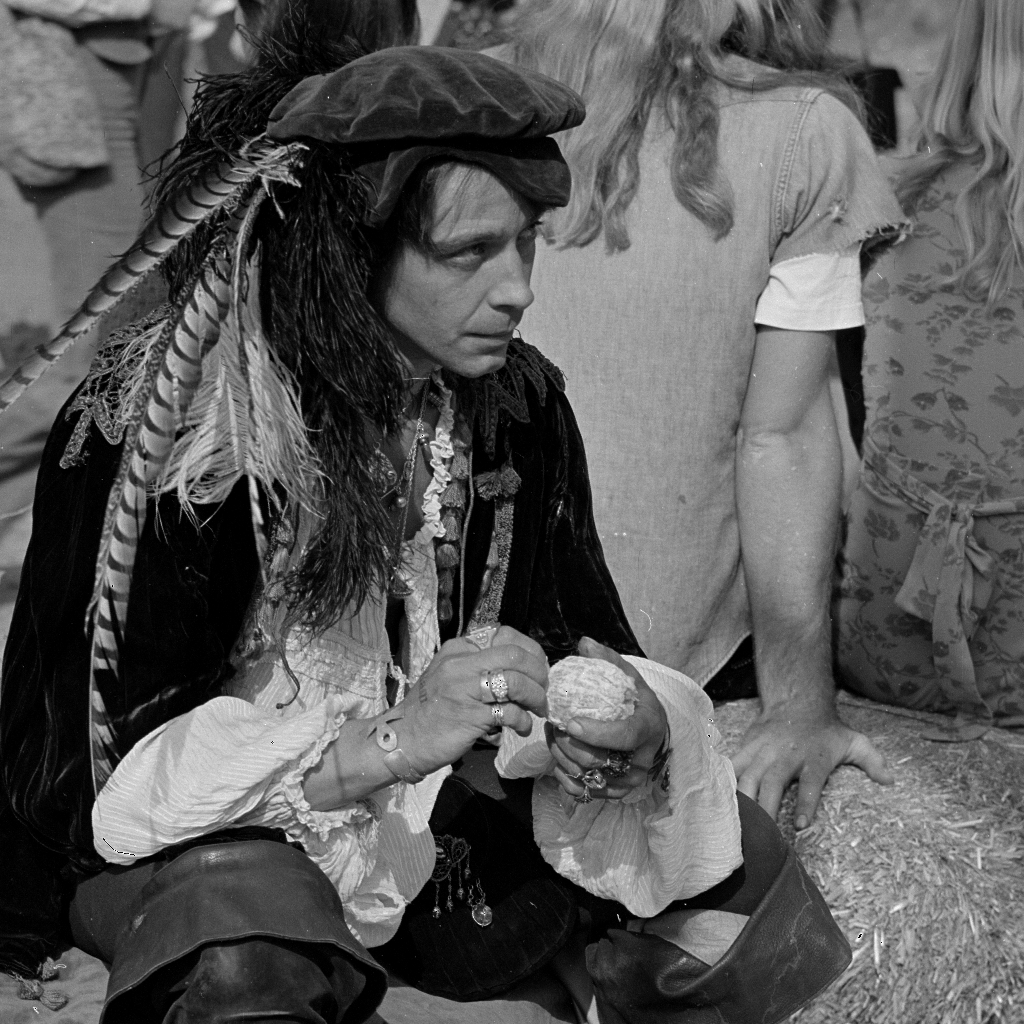}
        \caption{1024x024}
    \end{subfigure}
    
    \caption{Original and decrypted images at various resolutions}
    \label{fig:12iamges}
\end{figure*}

Upon completing the tests and analyzing the images, we observed that the decrypted images exhibit a high degree of similarity to the original images, as demonstrated in Figure~\ref{fig:12iamges}. To rigorously assess the effectiveness of our encryption methods in preserving image integrity and quality, we conducted extensive tests utilizing a range of image quality metrics. \par

We evaluated the Mean Squared Error (MSE), which measures the average squared difference between pixel values of the original and decrypted images. It is defined as the average of the squared differences between corresponding pixels of the two images~\cite{mse,  sara2019image}. The formula for MSE is $\text{MSE} = \frac{1}{mn} \sum_{i=1}^{m} \sum_{j=1}^{n} \left( I(i,j) - K(i,j) \right)^2$,
where \( I(i,j) \) and \( K(i,j) \) represent the pixel values at position \( (i,j) \) in the original and the decrypted images, respectively, and \( m \) and \( n \) are the number of rows and columns in the image.  A lower MSE indicates higher similarity between the original and decrypted images. 

We also assessed the Peak Signal-to-Noise Ratio (PSNR), which quantifies the reconstruction quality of the decrypted image. It is usually expressed in decibels (dB)~\cite{sara2019image}. The formula for PSNR is $\text{PSNR} = 10 \cdot \log_{10} \left( \frac{MAX_I^2}{\text{MSE}} \right),$
where \( MAX_I \) is the maximum possible pixel value of the image (e.g., \( MAX_I = 255 \) for an 8-bit image). A higher PSNR indicates better quality and higher similarity of a decrypted image to the original image.\par

As shown in Table \ref{tab:mse_psnr}, the low MSE and high PSNR values demonstrate that our encryption and decryption processes effectively preserve image quality. These results confirm that our caching-optimized encryption method not only maintains the integrity and visual fidelity of the processed images, but also significantly enhances encryption efficiency. In particular, we achieved an average MSE of 0.476 and an average PSNR of 51.30 dB. In the field of image processing, MSE values below 10 and PSNR values above 40 dB are typically considered to indicate high similarity between two images~\cite{mse,sara2019image}. Our results, therefore, indicate a high degree of similarity between the original and decrypted images after employing our FHE encryption method, ensuring minimal perceptual loss.

\begin{table}[h!]
\centering
\fontsize{8pt}{10pt}\selectfont
\begin{tabular}{>{\centering\arraybackslash}p{3cm}cc}
\toprule
\textbf{Image resolution} & \textbf{MSE} & \textbf{PSNR (dB)} \\ \midrule
8x8        & 0.453  & 51.56  \\
64x64      & 0.477  & 51.37  \\
128x128    & 0.480  & 51.71  \\
256x256    & 0.520  & 50.86  \\
512x512    & 0.479  & 51.31  \\
1024x1024  & 0.451  & 51.08  \\
\bottomrule
\end{tabular}
\caption{MSE and PSNR across different image resolutions}
\label{tab:mse_psnr}
\end{table}


\subsection{Real-world Applications}
\label{casestudy}
To demonstrate that \thename ~enables ciphertext-based
image processing in real-world applications, we will implement it across four distinct use cases: mean filtering, brightness enhancement, image matching, and watermarking.

\subsubsection{Mean Filtering}
Mean filtering is a fundamental image processing technique used for noise reduction and image smoothing. It replaces each pixel value with the average value of its neighboring pixels within a specified window. In this case study, we demonstrate the implementation of mean filtering on image ciphertext generated by our FHE method.\par
Given a pixel value at position (\textit{i}, \textit{j}) within an image \textit{I}, the mean filter operation in the plaintext domain is mathematically expressed as $I'_{(i,j)} = \frac{1}{n^2} \sum_{k=-\frac{n}{2}}^{\frac{n}{2}} \sum_{l=-\frac{n}{2}}^{\frac{n}{2}} I_{(i+k,j+l)}$,
where $n$ is the size of the filter window (e.g., $n = 3$ for a 3x3 filter), and $I'_{(i,j)}$ is the filtered pixel value~\cite{meanfilter,meanfilter2, meanfilter3, meanfilter4}.

\begin{figure}[htbp]
    \centering
    \begin{subfigure}[b]{0.2\textwidth}
        \includegraphics[width=\textwidth]{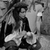}
        \caption{Original image}
    \end{subfigure}
    \hfill
    \begin{subfigure}[b]{0.2\textwidth}
        \includegraphics[width=\textwidth]{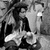}
        \caption{After mean filtering}
    \end{subfigure}
    \caption{Images before and after mean filtering, with (b) slightly brighter than (a) due to mean filtering}
    \label{fig: meanfilter}
\end{figure}

As shown in Figure \ref{fig: meanfilter}, we applied \thename ~to a 64$\times$64 pixel image, followed by a 3x3 mean filtering operation.  Frequency domain analysis confirmed that the decrypted image accurately reflected the application of the mean filter, verifying the correctness of our ciphertext-based homomorphic operations for image mean filtering. Specifically, the MSE between the original image and the filtered image is 218.13, with a PSNR of 24.74 dB. Given that a PSNR value between 20 dB and 30 dB indicates noticeable differences, these results highlight a significant loss of detail, as anticipated from the mean filtering.\par

Without using \thename, the  pixel-wise encryption time for this example was 84.29 seconds, whereas our full caching method reduced it to just 4.65 seconds, delivering an impressive 18-fold speedup.

\subsubsection{Brightness Enhancement}
Brightness enhancement is a common image processing technique used to improve image visibility.
We applied homomorphic operations to real images encrypted by \thename, performing brightness enhancement while the images remained encrypted. Specifically, we incremented each pixel value by 50, resulting in brighter images~\cite{enhance}. The image was then decrypted to visualize the processed effect. It is important to note that this brightness enhancement technique is effective only when encryption is applied at the pixel level.\par

\begin{figure}[htbp]
    \centering
    \begin{subfigure}[b]{0.2\textwidth}
        \includegraphics[width=\textwidth]{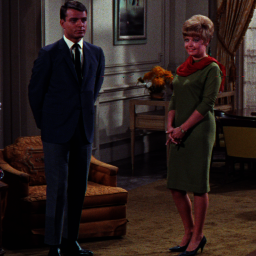}
        \caption{Original}
    \end{subfigure}
    \hfill
    \begin{subfigure}[b]{0.2\textwidth}
        \includegraphics[width=\textwidth]{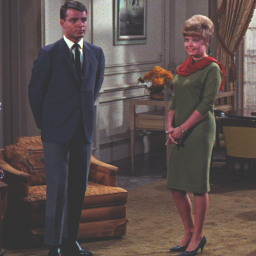}
        \caption{Processed}
    \end{subfigure}
    \caption{Brightness enhancement based on FHE-generated image ciphertext}
    \label{fig: enhance}
\end{figure}


Figure \ref{fig: enhance} illustrates the result of brightness enhancement applied to an encrypted image, showing a marked improvement in clarity. Our analysis confirms that the image processed based on ciphertext is indistinguishable from the image processed directly on plaintext.

In this example, \thename ~accelerated the image encryption by 4 times. For the 256$\times$256 image, we reduced the encryption time from 16.96 seconds to 4.57 seconds. 

\subsubsection{Image Matching}
Image matching is a fundamental task in computer vision, where the goal is to determine whether two images are similar. We're matching images using the L1 distance (also known as Manhattan distance) in a pixel-by-pixel manner and comparing the images directly based on their pixel values.
The L1 distance \( d_{L1} \) between two images \( I_1 \) and \( I_2 \) is $d_{L1}(I_1, I_2) = \sum_{i=1}^M \sum_{j=1}^N \left| I_1(i, j) - I_2(i, j) \right|$,
where \( I_1(i, j) \) and \( I_2(i, j) \) represent the pixel intensities at position \((i, j)\) in images \( I_1 \) and \( I_2 \), respectively~\cite{L1,L1dis}.

\begin{figure}[htbp]
    \centering
    \begin{subfigure}[b]{0.15\textwidth}
        \centering
        \includegraphics[width=\textwidth]{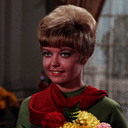}
        \caption{Image 1}
    \end{subfigure}
    \hfill
    \begin{subfigure}[b]{0.15\textwidth}
        \centering
        \includegraphics[width=\textwidth]{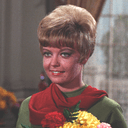}
        \caption{Image 2}
    \end{subfigure}
    \hfill
    \begin{subfigure}[b]{0.15\textwidth}
        \centering
        \includegraphics[width=\textwidth]{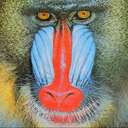}
        \caption{Image 3}
    \end{subfigure}
    \caption{Three images for matching}
    \label{figure:imagematch}
\end{figure}

Using three 128$\times$128 images for illustration, \added[id=SY]{as shown in Figure \ref{figure:imagematch}} , our specific operation involved applying \thename ~to encrypte them, followed by homomorphic subtraction and multiplication on the ciphertexts to compute the L1 distance in encrypted form. The user can then decrypt the distance using their key to determine which image in the database is most similar to the one they uploaded. This technique can be applied in facial recognition or medical image identification. \par

As shown in Figure \ref{figure:imagematch}, Image 2 is obtained by brightening Image 1 and Image 3 is a totally different one. As shown in Table \ref{table:distance}, the distances between the images are as follows: 11075.62 between Image 1 and Image 2; 21036.48 between Image 1 and Image 3; and 15133.89 between Image 2 and Image 3. The distance results are consistent with what we observe visually in Figure \ref{figure:imagematch}.

Without using \thename, the encryption time for this example was 826.19 seconds. By using \thename, it was reduced to 69.99 seconds. The performance has been improved by about 91.53\%, which is a significant enhancement.\par

\begin{table}[h]
\centering
\tiny
\resizebox{0.45\textwidth}{!}{
\begin{tabular}{c|ccc}
\hline
 & \textbf{Image 1} & \textbf{Image 2} & \textbf{Image 3} \\ 

\hline
\textbf{Image 1} & 0.00 & 11075.62 & 21036.48 \\ 
\textbf{Image 2} & 11075.62 & 0.00 & 15133.89 \\
\textbf{Image 3} & 21036.48 & 15133.89 & 0.00 \\ 
\hline
\end{tabular}
}
\caption{Comparison of distances between images}
\label{table:distance}
\end{table}


\subsubsection{Watermarking}
Watermarking images is crucial for ensuring the authenticity and ownership of digital content. In an era where digital media can be easily copied and distributed, embedding a watermark provides a layer of protection against unauthorized use and counterfeiting~\cite{fridrich1998image,shen2012single}. \par
We performed a watermarking operation on the encrypted images by adding a value of \( x \) to a specific pixel. The value of \( x \) is set by the user; for the convenience of the experiment, we chose a value of 5.0. This slight alteration creates a watermark within the encrypted image, ensuring that the watermark is embedded securely without revealing any underlying data. 
This approach is valuable for scenarios where image authenticity and traceability are critical. 

\begin{figure}[htbp]
    \centering
    \begin{subfigure}[b]{0.15\textwidth}
        \centering
        \includegraphics[width=\textwidth]{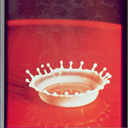}
        \caption{Original}
    \end{subfigure}
    \hfill
    \begin{subfigure}[b]{0.15\textwidth}
        \centering
        \includegraphics[width=\textwidth]{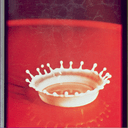}
        \caption{Watermarked}
    \end{subfigure}
    \hfill
    \begin{subfigure}[b]{0.15\textwidth}
        \centering
        \begin{tikzpicture}
            \node[anchor=south west,inner sep=0] (image) at (0,0) {\includegraphics[width=\textwidth]{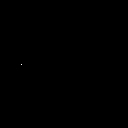}};
            \begin{scope}[x={(image.south east)},y={(image.north west)}]
                \draw[red, thick, ->] (0.5, 0.5) -- (0.3, 0.5);
            \end{scope}
        \end{tikzpicture}
        \caption{Comparison by bit}
    \end{subfigure}
    \caption{Images before and after adding watermark. The watermark is invisible but can be detected. The white area indicated by the red arrow in (c) represents the difference at this pixel between (a) and (b).}
    \label{fig: watermark}
\end{figure}

Adding a watermark at a specific pixel location can make the watermark invisible, as shown in Image (a) and (b) from Figure \ref{fig: watermark}. Image (c) from reflect the differences between (a) and (b): it is a binary image where the white areas represent the pixel locations with differences above the threshold. In this case, the thredshold is the value of 5.0 we chose before. This technique ensures that the watermark does not alter the visual appearance of the image, maintaining its aesthetic integrity.



In this example, without using \thename, the encryption time was 63.08 seconds. By using \thename, it was reduced to 4.8 seconds, achieving an 16-fold improvement in speed.

\subsection{Broader Applicability}
\label{generalizability}


To ensure that our approach is broadly applicable and robust across various image types and content, we conducted experiments using a diverse range of images, although not all could be displayed in this paper. To make sure the image selection captured a broad spectrum of variation, we calculated Structural Similarity Index (SSIM) values between images, quantifying their degrees of similarity and dissimilarity.

SSIM is a metric used to assess the similarity between two images, focusing on the preservation of structural information. Unlike MSE and PSNR, SSIM considers luminance, contrast, and structure in its assessment~\cite{sara2019image}. The formula for SSIM is $\text{SSIM}(I, K) = \frac{(2\mu_I \mu_K + C_1)(2\sigma_{IK} + C_2)}{(\mu_I^2 + \mu_K^2 + C_1)(\sigma_I^2 + \sigma_K^2 + C_2)}$,
where \( \mu_I \) and \( \mu_K \) are the mean values of images \( I \) and \( K \), \( \sigma_I^2 \) and \( \sigma_K^2 \) are the variances, \( \sigma_{IK} \) is the covariance between \( I \) and \( K \), and \( C_1 \) and \( C_2 \) are constants included to stabilize the division when the denominator is close to zero.

\begin{figure*}
    \centering
    \begin{subfigure}[b]{0.15\textwidth}
        \includegraphics[width=\textwidth]{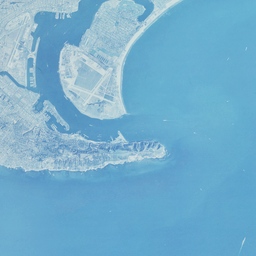}
        
    \end{subfigure}
    \hfill
    \begin{subfigure}[b]{0.15\textwidth}
        \includegraphics[width=\textwidth]{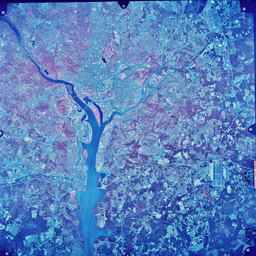}
       
    \end{subfigure}
    \hfill
    \begin{subfigure}[b]{0.15\textwidth}
        \includegraphics[width=\textwidth]{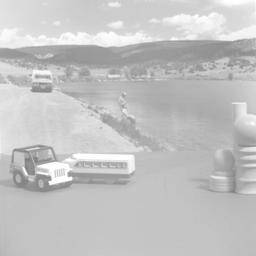}
      
    \end{subfigure}
    \hfill
    \begin{subfigure}[b]{0.15\textwidth}
        \includegraphics[width=\textwidth]{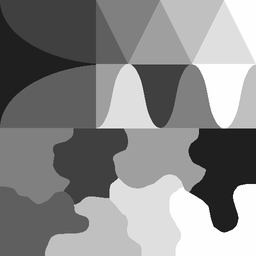}
        
    \end{subfigure}
    \hfill
    \begin{subfigure}[b]{0.15\textwidth}
        \includegraphics[width=\textwidth]{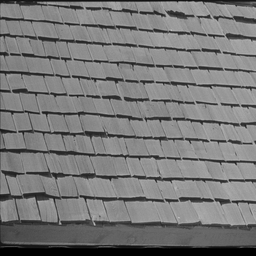}
       
    \end{subfigure}
     \hfill

    \vspace{1em}
    
    \begin{subfigure}[b]{0.15\textwidth}
        \includegraphics[width=\textwidth]{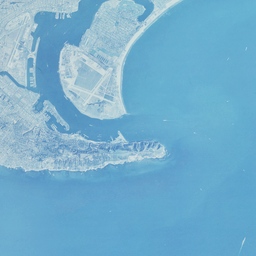}
        \caption{}
    \end{subfigure}
    \hfill
    \begin{subfigure}[b]{0.15\textwidth}
        \includegraphics[width=\textwidth]{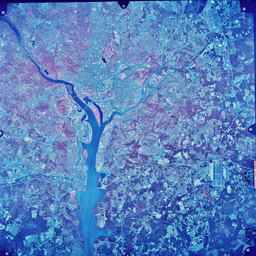}
        \caption{}
    \end{subfigure}
    \hfill
    \begin{subfigure}[b]{0.15\textwidth}
        \includegraphics[width=\textwidth]{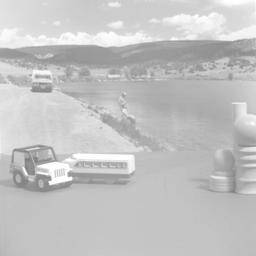}
        \caption{}
    \end{subfigure}
    \hfill
    \begin{subfigure}[b]{0.15\textwidth}
        \includegraphics[width=\textwidth]{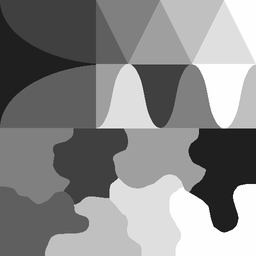}
        \caption{}
    \end{subfigure}
    \hfill
    \begin{subfigure}[b]{0.15\textwidth}
        \includegraphics[width=\textwidth]{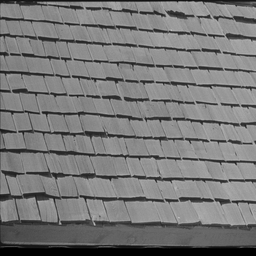}
        \caption{}
    \end{subfigure}
    \hfill

    \caption{Five examples out of 200 different types of pictures before encryption and after decryption. Images representing a range of scenarios: satellite images, toy vehicle, texture mosaic and wood shingle roof, to test the generalization of \thename.}
    \label{fig:image_set}
\end{figure*}

This guided our selection of 200 distinct images, each with dimensions of 256$\times$256 pixels, from the USC-SIPI~\cite{USCSIPI} database. Five representative examples from this image set are displayed in Figure~\ref{fig:image_set}. The top row showcases the five original images with distinct characteristics: three in grayscale and two in color. We applied the \thename\ operation to all the images. The bottom portion of Figure~\ref{fig:image_set} illustrates the images after undergoing encryption and subsequent decryption, with each restored to its original form. The identical appearance of the images before and after the process highlights the effectiveness and broader applicability of our approach.

\begin{figure}[!t]
    \centering
    \begin{subfigure}[b]{0.45\linewidth}
        \centering
        \includegraphics[width=\linewidth]{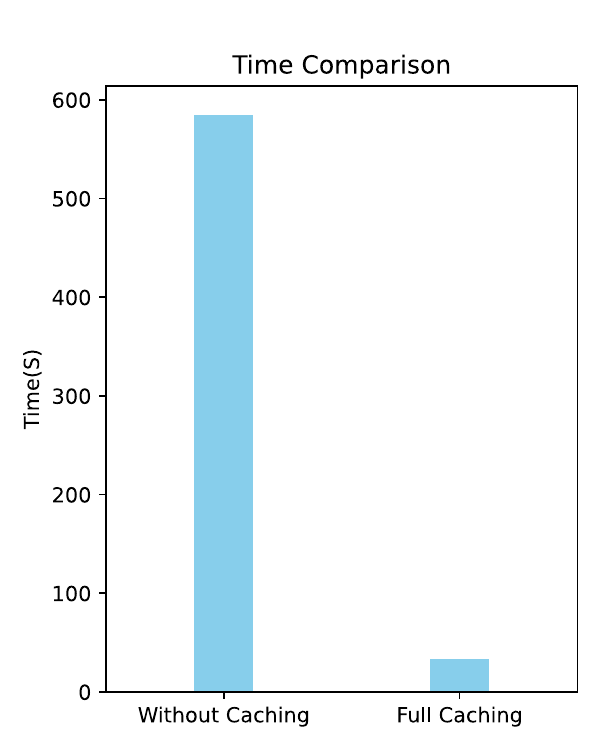}
    \end{subfigure}
    \hfill
    \begin{subfigure}[b]{0.45\linewidth}
        \centering
        \includegraphics[width=\linewidth]{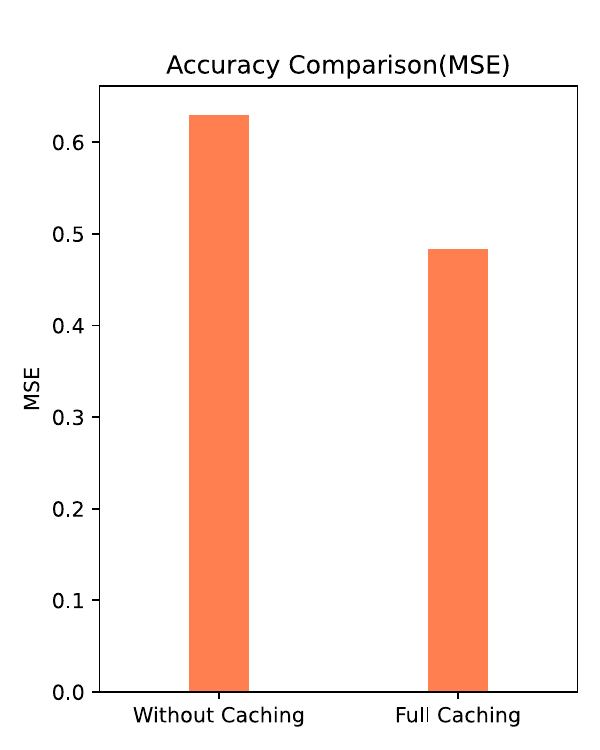}
    \end{subfigure}

    \caption{Comparison on time and accuracy for different types images}
    \label{generalization}
\end{figure}

Figure~\ref{generalization} presents the experimental results from processing images using both non-caching and full-caching approaches. The figure demonstrates a significant reduction in encryption time, with decrypted images closely matching the originals. Specifically, the average processing time decreased by a factor of 17.57, while the average MSE difference is 0.146. These results underscore the robustness and the broad applicability of our approach in practical scenarios~\cite{zhai2022accelerating}.

The SSIM values calculated between these images range from 0.0737 to 0.5640, reflecting a spectrum of similarity and diversity within the dataset. This variation in SSIM values corresponds to the differing structural characteristics among the images, illustrating the range of similarity and dissimilarity present. The variability in SSIM values highlights that the image set used in this study represents a wide range of image types, thereby reinforcing the broad applicability of our findings. This distribution of SSIM values suggests that the approach employed is robust across images with diverse structural content, thereby extending the applicability of our results to a broad array of real-world scenarios.


\section{Related Work}

This section reviews relevant work of FHE, starting with a general overview (Section \ref{relatedwork: FHE}), HE performance acceleration (Section \ref{relatedwork: acceleration}), followed by recent developments in the CKKS scheme (Section \ref{relatedwork: CKKS}), and ending with the CKKS application to image encryption (Section \ref{relatedwork: CKKS for images}).


\subsection{FHE}
\label{relatedwork: FHE}

Fully Homomorphic Encryption (FHE) represents a groundbreaking advancement in the field of cryptography, enabling computations on encrypted data without needing to decrypt it first. It was initially proposed by Rivest~\cite{Rivest1978} in 1978. Later, Gentry et al.~\cite{gentry2009} introduced pioneering work on FHE in 2009, which provided the first feasible construction of an FHE scheme, and since then there has been significant progress in making FHE more practical for real-world applications~\cite{blatt2020optimized} .

FHE schemes have evolved from theoretical constructs to more efficient implementations, addressing the challenges of high computational overhead and limited functionality. Various FHE schemes have been proposed, including BFV~\cite{FV2012,brakerski2012}, BGV~\cite{BGV2014},  CKKS~\cite{ckks2016}, TFHE~\cite{torusCircuit} and its optimizations, each tailored to specific types of computations and use cases. 

BFV and BGV both operate over integers, making them ideal for exact computations, with BFV generally offering faster performance for certain operations while BGV allows for deeper computational circuits before requiring bootstrapping. These distinctions have led to various optimization efforts. For instance, Halevi et al.~\cite{improveBFV2019} focused on enhancing the BFV scheme by optimizing decryption and homomorphic multiplication in the Residue Number System (RNS). Robin et al.~\cite{geelen2022basalisc} introduced BASALISC, the first to implement the BGV scheme with fully-packed bootstrapping, i.e., the noise removal capability necessary for arbitrary-depth computation. In contrast, CKKS is designed for approximate arithmetic on real and complex numbers, introducing controlled noise to enhance efficiency. This makes CKKS particularly well-suited for machine learning and signal processing tasks where minor approximation errors are acceptable~\cite{boemer2020mp2ml, han2023,imageclassification2}. While BFV and BGV provide precise results for integer arithmetic, CKKS trades some precision for improved performance and versatility in floating-point operations~\cite{Bossuat2021}. Additionally, TFHE, which was proposed by Chillotti et al.~\cite{torusCircuit}, introduces significant optimizations in bootstrapping and leveled homomorphic encryption on Boolean circuits, reducing computational overhead and improving the practicality of FHE for arithmetic functions. There have been numerous efforts to optimize TFHE, including works such as \cite{ccsTFHEAccelerator}, \cite{IEEETfhe}, and \cite{Guimares2024MOSFHETOS}. Unlike CKKS, schemes like BFV, BGV, and TFHE operate over integers rather than floating-point numbers, which are commonly used in image processing. The choice between these schemes depends on the specific requirements of the application, balancing factors such as data type, precision needs, computational depth, and overall performance considerations~\cite{HEStandard, LPR13, iliashenko2021faster}.

While the first group of schemes -- BFV, BGV, CKKS and TFHE -- fall under the category of FHE, allowing for arbitrary-depth computation, another line of research is Partial Homomorphic Encryption (PHE) methods. This paper does not focus on PHE and will therefore not discuss it in detail.

\subsection{HE Performance Acceleration}
\label{relatedwork: acceleration}

While FHE methods have made significant strides in security, they often encounter challenges like high computational costs and slow processing speeds in real-world applications. Recently, there have been many works focusing on the acceleration on homomophic encryption (HE). Agrawal et al.~\cite{agrawal2023mad,agrawal2023fab} propose memory-aware design (MAD) techniques to accelerate the bootstrapping operation of CKKS. Genise et al.~\cite{gadgetDecomposition} improve the runtime of the main bottleneck operation using ``gadget'' by 18x for 2 attributes and up to 289x for 16 attributes. Additionally, efforts in hardware acceleration have been investigated to speed up homomorphic operations. For instance, Di Matteo et al.~\cite{DiMatteo2023} propose hardware accelerator, presenting a configurable Number Theoretic Transform (NTT) unit for all the polynomial degrees available on the SEAL-Embedded. In addition to acceleration on FHE, PHE performance acceleration is also significantly improved. One of the state-of-the-art techniques is Rache~\cite{sigmod2023}, which introduces a radix-based parallel caching optimization for accelerating the performance of homomorphic encryption (HE) of outsourced databases in cloud computing. However, it is only available for accelerating PHE schemes like Paillier~\cite{paillier1999} and Symmetria~\cite{symmetria}, and was not designed for handling image data.




\subsection{CKKS}
\label{relatedwork: CKKS}

The CKKS scheme, introduced by Cheon, Kim, Kim, and Song in 2017\cite{ckks2016}, is a leveled homomorphic encryption scheme designed for approximate arithmetic. Unlike earlier schemes that focused on exact computations, CKKS allows for efficient encrypted computations with some tolerable error, which is often acceptable in real-world applications. This trade-off between accuracy and efficiency has made CKKS a popular choice for applications in privacy-preserving machine learning, encrypted databases, and secure cloud computing .

Recent work has focused on optimizing CKKS to reduce its computational overhead and enhance its practicality. Techniques such as bootstrapping, ciphertext packing, and key switching have been explored to improve the performance of CKKS-based systems. For example, Lee et al.~\cite{Lee2021, lee2022privacy} propose to compose the optimal approximate polynomial of the inverse sine function to the sine/cosine function to improve the precision of the bootstrapping. Cheon et al.~\cite{cheon2023} proposed Mult2 to perform ciphertext multiplication in the CKKS scheme with lower modulus consumption. Lee et al.~\cite{elasm} proposed a scheme to actively control the scale of a ciphertext, making the impact of noise on an error smaller. Bossuat et al.~\cite{Bossuat2021} optimize the key-switch procedure and propose a new technique for linear transformations. Chen et al.~\cite{packed2019} generalize the bootstrapping techniques for HE to obtain multi-key fully homomorphic encryption schemes.  Jain et al.~\cite{DBLP} proposed that the efficiency of ciphertext packing schemes, model optimization, and multi-threading strategies significantly impacts the throughput and latency during the inference process.

\subsection{CKKS for Images}
\label{relatedwork: CKKS for images}

Given the unique features of CKKS, its application in image encryption has been the subject of extensive research. Boemer et al.~\cite{CKKSImage2019} introduced nGraph-HE2, which optimizes ciphertext-plaintext addition and multiplication using the CKKS scheme. Similarly, Mertens et al.~\cite{imageCompression} proposed a practical image compression and processing pipeline tailored for CKKS, emphasizing FHE-friendliness.

Recently, there has been a growing interest in integrating CKKS with machine learning frameworks for images. For example, Jia et al.~\cite{jia2024} introduced Homomorphic Encryption Image Classification Evaluation (HEICE), a novel approach for secure image classification. Dimitris et al.~\cite{Dimitris} introduced a scheme combining federated learning and CKKS for neuroimaging analysis. Benaissa et al.~\cite{tenseal} presented TenSEAL, an open-source library for Privacy-Preserving Machine Learning based on CKKS.

Despite these advancements, challenges persist in making FHE schemes, including CKKS, more accessible and scalable. Issues such as key management, noise growth during computations, and the requirement for high-performance hardware remain active areas of research~\cite{Xiang2023}. Moreover, the trade-offs between security and efficiency—especially regarding bootstrapping and noise management—are critical considerations~\cite{cluster2024}. Nonetheless, the increasing focus on optimizing and applying CKKS in various domains, such as privacy-preserving machine learning~\cite{WANG2023103163} and secure data analytics, underscores its potential to become a cornerstone of secure computation in the era of big data and artificial intelligence.

\section{Conclusion}


In conclusion, this paper advances practical ciphertext-based image processing by proposing and implementing a pixel-level homomorphic encryption approach based on the CKKS scheme. Our approach employs three caching strategies -- radix-based caching, scanning-based caching, and full caching -- to pre-encrypt radix values and frequently occurring or all pixel values, thereby significantly reducing redundant encryption operations. Our evaluation across a diverse set of images demonstrates substantial improvements in encryption efficiency, with up to a 19-fold reduction in encryption time for large images, all while maintaining image quality. We further showcased the versatility and practicality of our approach by applying it to real-world image processing tasks, including mean filtering, brightness enhancement, image matching, and watermarking, all performed directly on ciphertexts. The results indicate notable performance gains, with encryption tasks becoming up to 91.53\% faster and achieving a 16-fold reduction in encryption time. Additionally, we validated that our approach is IND-CPA secure, benefiting from the integrated randomness.


{\footnotesize \bibliographystyle{acm}
\bibliography{sample}}


\end{document}